\documentstyle[prl,aps,twocolumn,epsf]{revtex}
\def\L{{\bf L}}
\def\F{{\bf F}}

\def\Ph{{\bf P}}
\def\K{{\cal K}}
\begin{document}
\twocolumn[\hsize\textwidth\columnwidth\hsize\csname@twocolumnfalse%
\endcsname
\rightline{McGill-99/36}
\title{Derivation of the Semi-circle Law from the Law of Corresponding States}
\author{C.P.~Burgess,$^{1}$ Rim Dib$^{1}$ and Brian P. Dolan$^{2,3}$\\
\vspace{3mm}
{\small
$^{1}$ Physics Department, McGill University,
3600 University Street, Montr\'eal, Qu\'ebec, Canada H3A 2T8.}\\
{\small
$^{2}$ Groupe de Physique Th\'eorique, Institut de Physique Nucl\'eaire,
F-91406 ORSAY Cedex, France}\\
\vspace{3mm}
{\small e-mail: cliff@physics.mcgill.ca, rimdib@physics.mcgill.ca,
bdolan@thphys.may.ie}}

\date{30th November 1999}

\maketitle

\begin{abstract}
We show that, for the transition between any two quantum Hall states, 
the semi-circle law and the existence of a duality symmetry follow
solely from the consistency of the law of corresponding states with 
the two-dimensional scaling 
flow. This puts these two effects on a sound 
theoretical footing, implying that both should hold 
exactly at zero temperature, independently of the details
of the  microscopic electron dynamics. 
This derivation also shows how the experimental evidence favours 
taking the two-dimensional flow seriously for the whole transition, 
and not just near the critical points.
\end{abstract}
\bigskip
\hfill{PACS nos: 73.40.Hm, 05.30.Fk, 02.20.-a} 
\bigskip

$^3${\small On leave from: 
Dept. of Mathematical Physics, 
National University of Ireland, Maynooth, Republic of Ireland.}

{\small Research funds for this work have been provided by
grants from NSERC (Canada), FCAR (Qu\'ebec), CNRS (France) and by Enterprise 
Ireland (Basic research grant no. SC/1998/739).}

\bigskip
]
\hoffset -10pt
Quantum Hall systems are remarkable for the high accuracy with which many
of their properties are known. 
Although an explanation based on general principles
(gauge invariance) has been
given by Laughlin \cite{LaughlinGI} for the precise quantization of 
the Hall conductivity on the integer Hall plateaux, a similar
understanding of the robustness of the other Hall features
does not yet exist. Our goal in
this paper is to derive some of these as general consequences of the
symmetries of the low-energy limit of these systems, independently
of the microscopic details.

The current understanding of the transport properties of quantum 
Hall systems is based on a very successful effective field theory
consisting of composite fermions \cite{Jaina}.
The symmetry
which we shall use in what follows was argued to be a property
of the effective theory, under certain circumstances, in a seminal
paper by Kivelson, Lee and Zhang (KLZ) \cite{KLZ}.
These authors argue that the effective theory satisfies
a law of corresponding states, which consist of the following 
correspondences between conductivities, in the long wavelength limit:
\begin{itemize}
\item
Landau Level Addition Transformation ({\bf L})
$$
\sigma_{xy}(\nu +1)\leftrightarrow \sigma_{xy}(\nu) + 1\quad
\sigma_{xx}(\nu +1)\leftrightarrow \sigma_{xx}(\nu);
$$
\item
Flux Attachment Transformation ({\bf F})
$$
\rho_{xy}\left({\nu\over 2\nu +1}\right)
\leftrightarrow \rho_{xy}(\nu) +2\quad
\rho_{xx}\left({\nu\over 2\nu +1}\right)
\leftrightarrow \rho_{xx}(\nu);
$$
\item
Particle-Hole Transformation ({\bf P})
$$
\sigma_{xy}(1-\nu)\leftrightarrow 1-\sigma_{xy}(\nu)\quad
\sigma_{xx}(1-\nu)\leftrightarrow \sigma_{xx}(\nu);
$$
\end{itemize}
where $\nu$ denotes the filling factor and we use units in 
which $e^2/h=1$. 

The arrows become equalities when the correspondence becomes
a symmetry, and the conditions for this to be the case are
discussed in \cite{KLZ} --- in particular this is expected to hold at zero
temperature. Taking repeated powers of $\L$, $\F$ and their
inverses generates an infinite order discrete group, which we shall
call the KLZ group and denote by $\K$, which is well known to
mathematicians (see e.g. \cite{Rankin} where it is
denoted by $\Gamma_U(2)$). The complete
group, including also the transformation $\Ph$, which is an outer 
auto-morphism of $\K$, was first proposed as relevant to the 
quantum Hall effect by L\"utken and Ross \cite{LutkenRossa}

The KLZ symmetry is most succinctly expressed by
writing 
the conductivity tensor 
as a single complex variable, 
$\sigma = \sigma_{xy}+ i\sigma_{xx}$ (with the resistivities therefore 
given by $\rho = -1/\sigma= -\rho_{xy}+i\rho_{xx}$). Since the ohmic 
resistance, $\sigma_{xx}$, must be positive, the physical region consists
only of the upper-half $\sigma$ plane. A general element of $\K$ 
can then be represented as $\gamma(\sigma)={a\sigma +b\over c\sigma +d}$
with integer coefficients, such that $c$ is even and $ad - bc=1$.
Note that 
$\K$ maps the upper half-complex conductivity plane into
itself. 

The whole upper-half conductivity plane can be obtained from 
the vertical strip above the semi-circle of unit diameter spanning 
$0$ and $1$ by repeated action of the KLZ group. This strip is termed 
the {\it fundamental domain} in the mathematical literature. 

The law of corresponding states can now be seen to imply that any quantum
Hall state can be obtained from any other state by the action of some
element of $\K$. Thus, for example, starting from $\sigma=1$ we obtain
the integer series $\sigma=n$ from $\L^{n-1}$, the Laughlin series
$\sigma={1\over 2m+1}$ from $\F^{m}$ and the Jain series $\sigma={p\over 2pm +1}$
from $\F^{m}\L^{p-1} $.
It has already been pointed out \cite{BD} that the KLZ group gives a selection
rule for quantum Hall transitions --- a transition between two
Hall plateaux with $\sigma_{xy}=p_1/q_1$  and $\sigma_{xy}=p_2/q_2$
is allowed only if $\vert p_1q_2-p_2q_1\vert=1$.
However it implies much more if we examine the
consequences for the $\beta$-functions of the theory.

Strong predictions can be made when KLZ symmetry is combined
with the scaling theory of disorder \cite{AALR}, as applied 
to quantum Hall systems \cite{Khem,Pruiskena,Huckestein}.
According to the scaling theory, conductances (so, in two dimensions,
also the conductivities) are the macroscopic measures of microscopic 
disorder, and so are the relevant variables whose renormalisation group
(RG) flow describes the system's long-wavelength evolution. The main 
tools for describing this flow are then the $\beta$ functions, which 
describe the scaling of 
$\sigma_{xx}$ and $\sigma_{xy}$. 

In this language the law of corresponding states becomes the requirement
that $\K$ (and $\Ph$) should commute with the RG flow 
\cite{Lutken,LutkenBurgess,BD}. 
The flow is described by a single complex 
$\beta$-function:
\begin{equation}
\beta(\sigma,\bar\sigma)={d\sigma\over dt}=
\beta_{xy}(\sigma_{xx},\sigma_{xy})+i\beta_{xx}(\sigma_{xx},\sigma_{xy}),
\end{equation}
and a simple calculation reveals that the flow commutes with the
symmetry if \cite{LutkenBurgess}
\begin{equation}
\label{wttwo}
\beta(\gamma(\sigma),\gamma(\bar\sigma))={d\gamma(\sigma)\over dt}
={\beta(\sigma,\bar\sigma)\over (c\sigma +d)^2} \; , 
\end{equation}
where the property $ad-bc=1$ has been used. 

We now describe some consequences which follow for the flow of
{\it any} $\beta$-function which satisfies eq.~(\ref{wttwo})
(and subject to a global requirement concerning flow topology,
as explained below) regardless of its detailed form.
Previous analyses have made further assumptions about the
functional form of $\beta$, \cite{LutkenBurgess,Crossover,Taniguchi,B+L,GMW}, 
but we shall avoid any such assumptions
here and simply follow the implications of particle-hole
symmetry. 

It is an immediate consequence of eq.~(\ref{wttwo}) that the
$\beta$-function must vanish at any point $\sigma_*$ (called a
fixed point) which is taken to itself --- {\it i.e.} $\gamma(\sigma_*) 
= \sigma_*$ -- by the action of a group element for which $c \sigma_* +d$ 
is neither zero nor infinite \cite{LutkenRossa,LutkenBurgess}.
The only such fixed points within the fundamental domain are the 
one at $\sigma_* = {1 \over 2}(1 + i)$ --- which is taken to itself by
$\gamma(\sigma) = (\sigma -1)/(2 \sigma -1)$ --- as well as 
$\sigma =0$ --- with $\gamma(\sigma) = -\sigma/(2 \sigma -1)$ --- 
and $\sigma = 1$ -- with $\gamma(\sigma) = (3\sigma -2)/(2 \sigma -1)$. 
$\beta$ must therefore vanish at these three points
(assuming it is finite). The consistency of KLZ
symmetry with the flow thereby predicts universal values for the
conductivity at the critical points, a possibility which was argued within
a more general context in ref.~\cite{Fisheretal}. 

KLZ symmetry also requires the $\beta$-function to 
vanish --- and to have precisely the same critical 
exponents --- at all of the images of the basic fixed points under 
the action of $\K$. There is indeed experimental evidence for this
equivalence of the critical exponents at different quantum Hall 
transitions (known as super-universality) \cite{WTPP,Engel}, a result
which had been also argued microscopically (neglecting electron 
self-interactions) \cite{microargs}.  

It is {\it not} an inescapable consequence of the KLZ symmetry that
there be no critical points other than those which are fixed points 
of $\K$. However there is no experimental evidence for any
other critical points in the quantum Hall effect, and so it would
not seem unreasonable to assume there is none. 
None of the following conclusions requires this assumption unless
explicitly stated.

Most of the above observations have already appeared in the literature
but we now go on to describe the two main results of this paper,
which have not been derived from general principles before.

{\sl 1. The Semi-circle Law:}
We now show that particle-hole symmetry, together with $\K$ invariance,
implies the semi-circle law.
The proof of this argument hinges on the existence of a unique 
and well-known function, $f(\sigma)$, which has the following two 
crucial properties \cite{Rankin}: 
1. It provides a one-to-one map from the fundamental 
domain to the complex plane (including the point at infinity); 
2. It is invariant under the action of the symmetry group $\K$, 
$f(\gamma(\sigma))=f(\sigma)$. Define a $\beta$-function for $f$,
\begin{equation}
B(f,\bar f):= {df \over dt}={df\over d\sigma}{d\sigma\over dt}={df\over
d\sigma}\beta(\sigma,\bar\sigma).
\end{equation}
We can conclude that $B(f,\bar f)$ is invariant with respect to
$\K$ --- since $f$ is already
invariant, $\K$-invariance imposes no further restrictions on the
function $B(f,\bar f)$.

Now impose particle-hole
symmetry, $\Ph$. To determine the consequences for $B(f,\bar f)$
due to $\Ph$ we use the following explicit expression \cite{Rankin}
for $f$ in terms of Jacobi $\vartheta$-functions (a very clear 
description of these classical functions is given in \cite{WW}):

\begin{equation}
f(\sigma)=-{\vartheta_3^4\vartheta_4^4\over\vartheta_2^8}=
-{1\over 256q^2}\prod_{n=1}^\infty
{\bigl(1-q^{4n-2}\bigr)^8\over \bigl(1+q^{2n}\bigr)^{16}} ,
\end{equation}
where $q=e^{i\pi\sigma}$. Since the action of particle-hole symmetry 
on $q$ is $\Ph: q\rightarrow -\bar q$, it is clear from the definition of 
$f$ that $\Ph: f = f(-\bar q) = \overline{f(q)}$. Thus particle-hole 
symmetry implies that $B(f,\bar f)$ must be invariant under the
interchange of $f$ and $\bar f$. So 
this implies that
\begin{equation}
\label{keyeq}
{d f\over dt}={B(f,\bar f)},\quad
{d\bar f\over dt}=\overline{B(f,\bar f)}=B(\bar f,f).
\end{equation}

Now suppose we start our RG flow
from a value of $\sigma$ for which $f$ is real.  Equation (\ref{keyeq})
states that $B$ is real when evaluated at this point, and hence 
${df \over dt}$ must be real. Repeating this argument point-by-point
along the flow line we see that particle-hole symmetry implies $f$ 
cannot develop an imaginary part if it doesn't start with one. We 
conclude that {\it curves on which $f$ is real are integral
curves of any renormalisation group flow which commutes with both
$\K$ and $\Ph$}! 

The curves along which $f$ is real are easily found, and 
for the fundamental domain consist of the curves defining
the boundaries, plus the vertical line $\sigma = {1\over 2}
+ i w$, $w \ge {1\over 2}$. 
$f$ is real along the vertical lines $\sigma={n\over 2}+iw$ (with 
$n$ integral) because it is an even function of $q$, and $q$ is 
real or pure imaginary when evaluated along these vertical lines.  
To see that $f$ is also real on the semi-circular arch spanning 
$0$ and $1$ requires the following classical facts about the 
$\vartheta$-functions (see e.g. \cite{WW}, page 475):
\begin{eqnarray}
\vartheta_2\left(-{1/\sigma}\right)=
\sqrt{-i\sigma}\vartheta_4(\sigma),&\quad&
\vartheta_3\left(-{1/\sigma}\right)=
\sqrt{-i\sigma}\vartheta_3(\sigma)\\
\vartheta_4\left(-{1/\sigma}\right)&=&
\sqrt{-i\sigma}\vartheta_2(\sigma).
\end{eqnarray}
The factors of $\sqrt{-i\sigma}$ here cancel when we take ratios to 
form $f$ and so $f(-1/\sigma)=-{\vartheta_3^4(\sigma)\vartheta_2^4(\sigma) 
\over \vartheta_4^8(\sigma)}$. Now $\sigma\rightarrow -1/\sigma$ sends 
the vertical line $\sigma=1+iw$, $0<w<\infty$, onto the semi-circle 
spanning $1$ and $0$. Since $\vartheta^4_a$ (${a=2,3,4}$) are all real 
on said vertical line (see e.g. \cite{WW}), 
$f$ must be real on said semi-circle, the latter 
is then perforce an integral curve of the renormalisation group flow.

The complete set of integral curves is then obtained by mapping
the above curves around the complex plane using ${\K}$,
and this is how figure 1 is generated.
Points where trajectories cross are fixed points of the renormalisation group 
flow, and the fixed point at $\sigma_*={1+i\over 2}$ is evident. 
The direction of the flow lines is uniquely determined if we assume that 
there are no other fixed points, the
Hall plateau are attractive fixed points of the flow, and that the flow comes
downwards vertically from $\sigma=i\infty$. The line segment
$\sigma=-{1\over2}+iw$, ${1\over 2}<w<\infty$ is mapped onto
$\sigma={1\over2}+iw$, $0<w<{1\over 2}$ by $\F$ --- the latter line must
therefore flow {\it upwards} towards ${1+i\over 2}$ if the former flows
downwards towards ${-1+i\over 2}$. Assuming that $0$ and $1$ are attractive
fixed points then determines the flow direction
as indicated by the arrows in figure 1.
Notice we are inevitably led to the existence of the semi-circles 
linking $0$ to $1/2$ in figure 1. 

It is a general property that the KLZ group takes semi-circles centred
on the real axis to other semi-circles also centred on the real axis
(including the degenerate case of infinitely large semi-circles, which
are vertical lines parallel to the imaginary axis). It follows 
that the flow between any two Hall plateaux must be along a semi-circle,
centred on the real axis, which is the image of the basic semi-circle 
connecting 0 and 1. 

In this way we obtain a robust derivation of the semi-circle law, 
which states that the conductivities move along such semi-circles 
in the conductivity plane during transitions between Hall plateaux. 
(Since the relation $\rho = -1/\sigma$ maps, for example, semi-circles 
having one end at $\sigma = 0$ into vertical lines in the $\rho$ plane, 
the corresponding statement in the resistivity plane 
is that for transitions from Hall fluids to the Hall insulator,
the flow is along lines of constant $\rho_{xy}$.)
Although the semi-circle law was 
proposed in \cite{Ruzin} on the basis of a particular microscopic 
model, we see here that it holds more generally than does its original
derivation. Any model compatible with the symmetry of the law of 
corresponding states must reproduce it. Experimentally, the
law is also well supported \cite{Hilkeetal}. 

\vtop{
\includegraphics{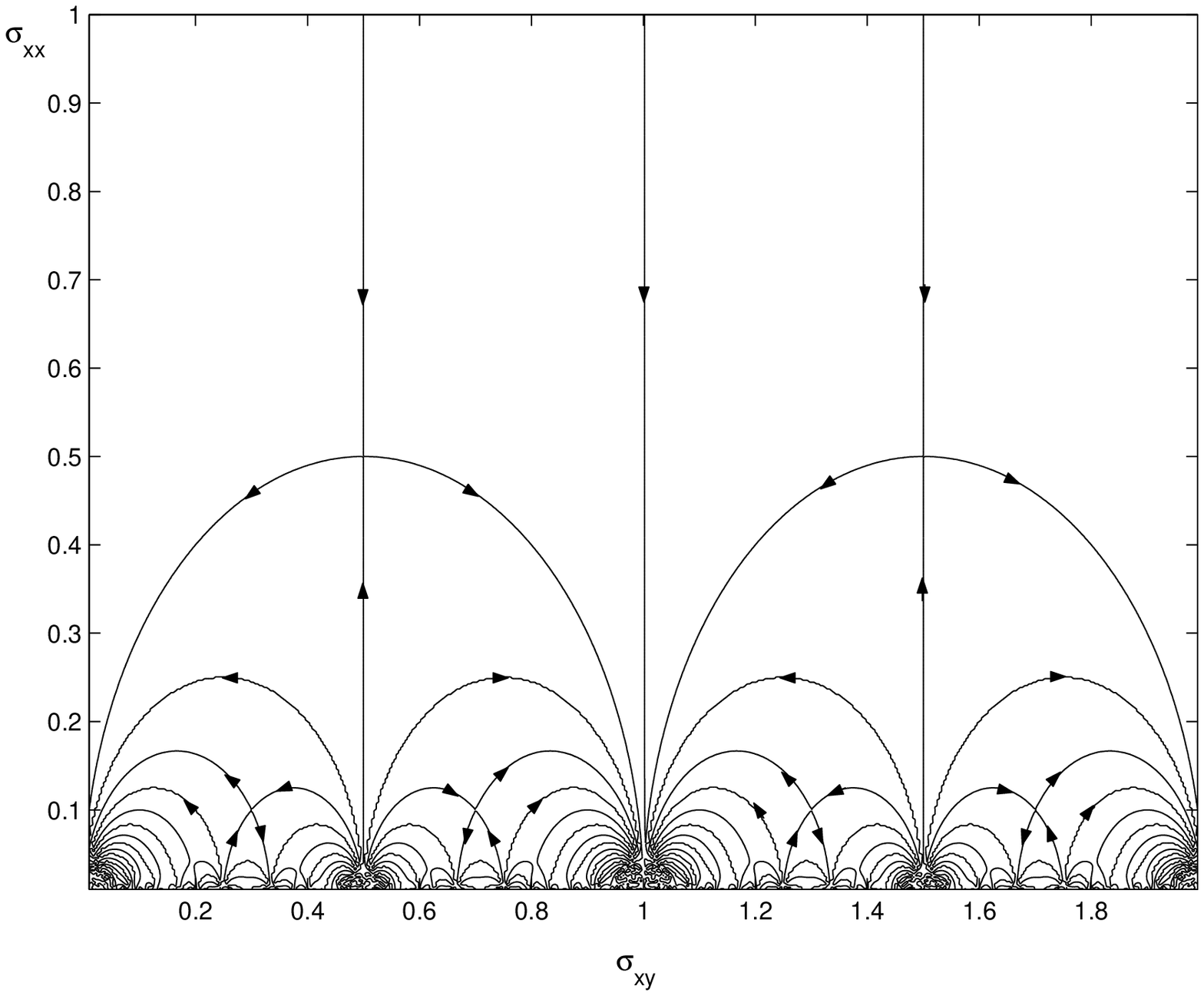}
\vskip 5.5cm
\centerline{Figure 1}}

\smallskip\noindent
{\sl 2. Duality:}
A second experimentally striking result which follows quite
generally from the symmetry version of the Law of Corresponding States
is the existence of a duality symmetry relating the conductivities 
of the flow on either side of the critical point as one flows between 
{\it any} two allowed Hall plateaux, or between Hall plateaux and the 
Hall insulator. 

Since all flows are related by a symmetry to the basic semi-circle
running between 0 and 1, we derive the duality symmetry for this
semi-circle in detail. A convenient parameterization of the semi-circle 
from 0 to 1 is $\sigma= {1\over 2} +{1\over 2}({1-w^2 +2iw\over 1+w^2})$,
with $0 < w < \infty$. 
The key observation is that this curve is reflected into itself 
about the vertical line $\Re\sigma = {1\over 2}$ by $\Ph$ -- as 
well as by $\gamma = {(\sigma - 1)\over (2 \sigma - 1)} \in \K$. In terms of
the parameter $w$ this becomes $w\rightarrow 1/w$, and since
the semi-circle transformed to the $\rho$ plane is $\rho = -1 + iw$,
this is recognizable as the $\rho_{xx}\rightarrow 1/\rho_{xx}$
duality which has been observed \cite{STSCSS} in the transition to 
the Hall insulator from the $\nu = 1$ integer Hall state.  

The extension to other transitions follows from the action of
$\K$. For $\nu:{p_1\over q_1}\rightarrow{p_2 \over q_2}$ with 
$p_2q_1-p_1q_2=1$, where the transition is along the curve
$\rho={-(q_2p_2+w^2q_1p_1)+iw\over (p_2^2 + w^2 p_1^2)}$, the
duality is again given by $w\rightarrow 1/w$. As specialized
to transitions to the Hall insulator from the Laughlin sequence,
$\nu:{1\over 2n+1}\rightarrow 0$, the flow is the vertical line 
$\rho=-(2n+1)+iw$ in the resistivity plane 
and so the duality $w\rightarrow 1/w$ again implies the inversion 
$\rho_{xx}\rightarrow 1/\rho_{xx}$ about the critical point.

In conclusion, we wish to emphasise two points.
First, the assumption that the law of corresponding states
holds as a symmetry at low temperatures leads to an infinite
order discrete symmetry group for the quantum Hall effect --- called
the KLZ group here. This group acts on the upper-half complex
conductivity plane. If this is to be a symmetry its action must
commute with the renormalisation group flow of the system
and fixed points of the group action must be fixed
points of the flow. Three kinds of fixed points are predicted in this
way: attractive fixed points with $\sigma_{xx} = 0$ (which are
images under the group of the basic ones at $\sigma = 0$ or 1
and all have odd denominator)
describing the quantum Hall fluids and the Hall insulator;
repulsive fixed points with $\sigma_{xx}=0$ (which are images of
$\sigma=1/2$ and all have even denominator) and
saddle points (which are images under the group of the basic one
at $\sigma = {1\over 2}(1+i)$) describing the critical points in
the transitions between the various quantum Hall states.  
By organizing the critical points  of the system {\it via} a
(infinite order) discrete symmetry, the KLZ group furnishes a
fascinating generalisation of the Kramers-Wannier ${\bf Z}_2$ duality of the 
Ising model. 

Particle-hole symmetry places further restrictions
on the $\beta$-function and dictates that the form of the
RG flow between Hall plateaux, and between the plateaux and the
Hall insulator, to described by semi-circle law. It also inescapably
predicts the general existence of a duality symmetry for all Hall
transitions, which reduces to the observed $\rho_{xx}\rightarrow1/\rho_{xx}$ 
duality for Laughlin-sequence/Hall-insulator transitions. 

A point that must be addressed here is that the experimental data
do not always reproduce a fixed point exactly at $\sigma_{xx}=1/2$
for integer transitions. For example in \cite{Hilkeetal}
the critical point in the $1\rightarrow 0$ transition
is definitely not identified with ${1+i\over 2}$.
This is therefore incompatible with the law of corresponding states.
However, it is notoriously difficult to extract the longitudinal
resistivity (and hence conductivity) from the longitudinal
resistance (this issue is discussed, for example, on page 52 in Cage's
article in \cite{PrangeGirvin}) --- the relationship between these two quantities
is plagued with geometrical ambiguities, unlike the transverse
resistivity. Since a Hall-insulator transition ${1\over 2q+1}\rightarrow 0$
in the $\sigma$
plane corresponds to a vertical line
$\rho=-(2q+1)+iw$, $w>0$, in the
resistivity plane, rescaling the imaginary part
of $\rho$ does not affect the semi-circle law in the $\sigma$ plane
for these transitions. However it {\it does} move
the critical point. One interpretation of the experiment \cite{Hilkeetal}
is that the experimentally determined longitudinal resistivity
is not the same as the true microscopic longitudinal resistivity,
but a constant multiple of it (a factor of $1.7$ in this particular experiment).

We thank C.A. L\"utken for helpful discussions. C.B. is grateful to
the University of Barcelona, and B.D. to the IPN, Orsay and CNRS,
for their generous support and kind hospitality while part of this 
work was carried out.


\begin{thebibliography}{99}
%
\bibitem{LaughlinGI} R. Laughlin, Phys. Rev. {\bf B23}, 5632 (1981).
%
\bibitem{Jaina}J.K.~Jain, Phys. Rev. Lett. {\bf 63}, 199 (1989);
Adv. Phys. {\bf 41}, 105 (1992).
%
\bibitem{KLZ} S.~Kivelson, D-H.~Lee and S-C.~Zhang, Phys. Rev. 
{\bf B46}, 2223 (1992).
%
\bibitem{Rankin} R.A.~Rankin, {\sl Modular Forms and Functions}, 
(Cambridge University Press, 1977).
%
\bibitem{LutkenRossa} C.A.~L\"utken and G.G.~Ross, 
Phys. Rev. {\bf B45}, 11837 (1992); Phys. Rev. {\bf B48}, 2500 (1993).
\bibitem{AALR}
E. Abrahams, P.W. Anderson, D.C. Licciardello and T.V. Ramakrishnan,
Phys. Rev. Lett. {\bf 42}, 673 (1979).
%
\bibitem{Khem} D.E.~Khmel'nitskii, Pis'ma Zh. Eksp. Teor. Fiz {\bf 38}, 
454 (1983) (JETP Lett. {\bf 38}, 552 (1983))
%
\bibitem{Pruiskena} A.M.M.~Pruisken, Phys. Rev. Lett. {\bf 61}, 1297 (1988)
%
\bibitem{Huckestein}
B. Huckelstein, Rev. Mod. Phys. {\bf 67}, 357 (1995);
S.L. Sondhi, S.M. Girvin, J.P. Carini and D. Shahar, Rev. Mod. Phys.
{\bf 69}, 315 (1997).
%
\bibitem{Lutken} C.A.~L\"utken, Nuc. Phys. {\bf B396}, 670 (1993).
%
\bibitem{LutkenBurgess} C.P.~Burgess and C.A.~L\"utken, 
Nuc. Phys. {\bf B500}, 367 (1997) (cond-mat/9611118).
%
\bibitem{BD} B.P.~Dolan, J. Phys. {\bf A32}, L243 (1999) (cond-mat/9805171).
%
\bibitem{Crossover} B.P.~Dolan, Nuc. Phys. {\bf 460B [FS]}, 297 (1999)
(cond-mat/9809294).
%
\bibitem{Taniguchi} N.~Taniguchi, 
{\sl Nonperturbative Renormalisation Group Function for 
Quantum Hall Plateau Transitions Imposed by Global Symmetries}, 
(cond-mat/9810334).
%
\bibitem{B+L} C.P.~Burgess and C.A.~L\"utken, Phys. Lett. {\bf B451}, 365
(1999) (hep-th/9812396).
%
\bibitem{GMW} Y.~Georgelin, T.~Masson and J-C.~Wallet, 
{\sl $\Gamma(2)$ Modular Symmetry,
Renormalization Group Flow and the Quantum Hall Effect}, (cond-mat/9906193).
%
\bibitem{Fisheretal}
M.P.A. Fisher, G. Grinstein and S.M. Girvin, Phys. Rev. Lett.
{\bf 64}, 587 (1990);
M.-C. Cha, M.P.A. Fisher, S.M. Girvin, M. Wallin and A.P. Young,
Phys. Rev. {\bf B44}, 6883 (1991).
%
\bibitem{WTPP}
H.P. Wei, D.C. Tsui, M.A. Paalanen and A.M.M. Pruisken,
Phys. Rev. Lett. {\bf 61}, 1294 (1988).
%
\bibitem{Engel}
L. Engel {\it et.al.}, Surf. Sci. {\bf 229}, 13 (1990).
%
\bibitem{microargs}
D.E.~Khmel'nitskii in ref.~[7]; A.M.M. Pruisken, in refs.~[8] and [25];
R.B. Laughlin, Phys. Rev. Lett. {\bf 52}, 2304 (1984);
%
\bibitem{WW} E.T.~Whittaker and G.N.~Watson, {\sl A Course of Modern 
Analysis}, Cambridge University Press, 1940).
%
\bibitem{Ruzin}  A.M.~Dykhne and I.M.~Ruzin, Phys. Rev. {\bf B50}, 
2369 (1994); 
I.~Ruzin and S.~Feng, Phys. Rev. Lett. {\bf 74}, 154 (1995).
%
\bibitem{Hilkeetal} M.~Hilke et al. {\sl Semicircle: an exact relation in the 
Integer and Fractional Quantum Hall Effect}, (cond-mat/9810217).
%
\bibitem{STSCSS} D.~Shahar, D.C.~Tsui, M.~Shayegan, J.E.~Cunningham,
E.~Shimshoni and S.L.~Sondhi, Solid State Comm. {\bf 102}, 817
(1997) (cond-mat/9607127)
D.~Shahar, D.C.~Tsui, M.~Shayegan, E.~Shimshoni and S.L.~Sondhi,  
Science {\bf 274}, 589 (1996) (cond-mat/9510113).
\bibitem{PrangeGirvin} R.E.~Prange and S.M.~Girvin (ed.), 
{\sl The Quantum Hall Effect}, (1987) Springer.
%
\end{thebibliography}
\end{document}